%% file: ms.tex
\documentclass[12pt,preprint]{aastex62}  

\usepackage{amssymb}
\usepackage{times}
\usepackage{verbatim}
\usepackage{color}

\usepackage{xspace}
\input{local_defs}

\shorttitle{High Resolution X-ray Spectra of NGC 3603}
\shortauthors{Huenemoerder et al.}

\begin{document}

\title{Winds of Massive Stars: High Resolution X-ray Spectra of Stars
  in NGC 3603}

\author[0000-0002-3860-6230]{David P.\ Huenemoerder}
\affiliation{Massachusetts Institute of Technology \\
77 Massachusetts Ave.\\
Cambridge, MA 02139, USA}

\author{Norbert S.\ Schulz}
\affiliation{Massachusetts Institute of Technology \\
77 Massachusetts Ave.\\
Cambridge, MA 02139, USA}


\author[0000-0003-3298-7455]{Joy S.\ Nichols}
\affiliation{Harvard-Smithsonian Center for Astrophysics\\
60 Garden St.\\
Cambridge, MA 02138, USA}




\begin{abstract}

  The cluster NGC~3603 hosts some of the most massive stars in the
  Galaxy.  With a modest 50~ks exposure with the Chandra High Energy
  Grating Spectrometer, we have resolved emission lines in spectra of
  several of the brightest cluster members which are of WNh and O
  spectral types.  This observation provides our first definitive
  high-resolution spectra of such stars in this nearby starburst
  region.  The stars studied have broadened X-ray emission lines, some
  with blue-shifted centroids, and are characteristic of massive
  stellar winds with terminal velocities around $2000$--$3000\kms$.
  X-ray luminosities and plasma temperatures are very high for both
  the WNh and O stars studied.  We conclude that their X-rays are
  likely the result of colliding winds.

\end{abstract}




\section{Introduction}\label{sec:intro}

The starburst phenomenon is an influential process in the evolution of
stellar systems.  The formation of massive stars can have a profound
effect on their environment through ionizing radiation, kinetic energy
of massive winds, and ultimately supernovae explosions.  \ngc is a
relatively nearby galactic starburst region, close enough that
individual stars can be observed in detail.  It is well recognized as
an important cluster in relation to the similar but more distant R136a
association, the much more active Arches cluster, and the distant
powerhouses in starburst galaxies.  

Mass loss rates and compositions can be determined through
high-resolution spectroscopy.  In X-rays, we can characterize emission
mechanisms, discriminating between wind-shocks, magnetic confinement,
or colliding winds.  X-ray emission line strengths and shapes are key
diagnostics of wind structure.

Some early \chan observations of hot stars challenged the canonical
wind-shock models \citep{Gagne:Oksala:al:2005,schulz2003} when some
massive stars, such as $\theta^1\,\mathrm{Ori\, C}$ were found to have
narrow lines and high temperatures. \citet{Gagne:Oksala:al:2005}
successfully applied a magnetically confined wind model to the
$\theta^1\,\mathrm{Ori\, C}$ spectra, and this is now largely accepted
as an explanation for narrow lines in hot stars.  Other stars, such as
$\zeta\,\mathrm{Pup}$ do show wind-broadened profiles
\citep{Cassinelli:Miller:al:2001, kahn2001, Kramer:Cohen:al:2003}.
The wind-shock models for single stars do not predict the high plasma
temperatures seen in some stars.  In binary systems, colliding winds
can lead to high temperatures since the temperature is proportional to
the square of the wind velocity \citep{Luo:McCray:al:1990,
  Stevens:Blondin:al:1992}.

\ngc is the closest starburst cluster at a distance of $\sim7.6\kpc$
and an age of about $1$--$3\,\mathrm{Myr}$
\citep{Melena:Massey:al:2008, Stolte:Brandner:al:2006,
  Hofmann:Seggewiss:al:1995}.  \citet{Stolte:Brandner:al:2006}
compared \ngc's central Young Cluster (YC) to other clusters and note
that it has a similar core radius to the Orion Nebular Cluster (ONC)
of about $0.2\,\mathrm{pc}$, and even a similar stellar number density
of about $2\times10^4\,\mathrm{pc\mthree}$, but its mass density of
$1\times10^5\,\mathrm{M_\odot\,pc\mthree}$ is 5 times that of the ONC;
the \ngc YC core mass equals the entire mass of the ONC.  The Arches
cluster has somewhat higher mass and mass density than \ngc YC; R136
is very similar to \ngc in many ways \citep{Moffat:Drissen:al:1994,
  Stolte:Brandner:al:2006}. Hence, the \ngc Young Cluster represents
and important object to study among others as, in the words of
\citet{Stolte:Brandner:al:2006}, it {\em ``provides a resolved
  template for extragalactic star-forming regions.''}

The stars in the core of \ngc are collectively known as \hdn and also
WR~43.  
\citet{Moffat:Seggewiss:al:1985} first identified the central
unresolved object as multiple WN-type
stars. \citet{Hofmann:Seggewiss:al:1995} resolved the stellar core of
\ngc into 28 stars using speckle interferometry.  These stars reside
in a central $6\times6\,\mathrm{arcsec^2}$ field (at $7.6\kpc$,
$6\,\mathrm{arcsec}$ corresponds to $0.22\pc$).  They identified
several of the components as late WN or Of
stars. \citet{Crowther:Dessart:1998} conducted a spectroscopic
analysis of the core stars in \ngc and determined their fundamental
properties. \hdn was resolved into several different components, and
three of them, A1, B, and C, have been classified as WN6h+abs types
--- nitrogen-rich Wolf-Rayet stars with substantial abundance of
hydrogen; they have not evolved past hydrogen-core burning.
\citet{Melena:Massey:al:2008} obtained spectra of \ngc stars,
determining characteristics for 16 additional objects; they also
compared to many of the previous results.  The three WN stars' winds
contribute about 65\% of the cluster kinetic energy, which is more
than that from cluster's 20--30 other O-stars combined
\citep{Crowther:Dessart:1998}.

In the \acis image data analyzed by \citet{Moffat:Corcoran:al:2002},
source C was the brightest in X-rays.  \hdn-C is a WN6h+abs type, with
$L/L_\odot\sim10^6$, $M/M_\odot\sim62$,
$\dot{M}\sim10\mfour\,\mathrm{M_\odot\,yr\mone}$, and a terminal wind
velocity of $2500\,\mathrm{km\,s\mone}$ \citep{Crowther:Dessart:1998}.
It is also a single-lined spectroscopic binary
\citep{Schnurr:Casoli:al:2008} with a period of $8.9\,\mathrm{days}$
and a velocity amplitude of $200\,\mathrm{km\,s\mone}$.

The Wolf-Rayet (WR) stellar class contains some of the most massive
and luminous stars.  Their dense, high velocity stellar winds, and
ultimate supernova explosion significantly affect the composition and
dynamics of the interstellar medium.  They are important in galactic
feedback and can alter the environment in their host star cluster.
There is some evidence that the X-ray production in WR stars is
perhaps different from known O-star mechanisms of wind-shocks, which
are thought to occur near the stellar photosphere in the wind
acceleration zone, or from magnetically confined winds, in which
strong magnetic fields constrain the winds.  High-resolution X-ray
spectra of WR$\,6$ (EZ~CMa), from \xmm-Newton \rgs and from
\chan/\hetg spectrometers showed that the X-rays are generated far out
in the wind \citep{Oskinova:al:2012, Huenemoerder:al:2015}.  The X-ray
emission mechanism is not known. Hence, more high resolution spectra
of very massive stars are required to further study winds in this
regime.  The central cluster of NGC~3603 provides us with an
opportunity to study in detail some of the most massive stars in the
Galaxy.


%

\section{Observations and Calibration}\label{sec:obscal}

Using the \chan/\hetg spectrometer \citep{HETG:2005}, we observed
NGC~3603 in 2011 for $47\ks$ (observation ID 13266). The \hetgs
spectra cover the range from about $1$--$30\mang$, as dispersed by two
types of grating facets, the High Energy Grating (\heg) and the Medium
Energy Grating (\meg), with resolving powers ranging from 100 to 1000
and approximately constant full-width-half-maxima ($FWHM$) of
$12\mmang$ for \heg and $23\mmang$ for \meg.  Our observation was
designed as a ``snapshot'', just long enough to characterize the
strongest emission lines of the brightest members, based on the
low-resolution \acis observations.

The \chan data were reprocessed with standard Chandra Interactive
Analysis of Observations (\ciao) programs \citep{CIAO:2006} to apply
{ calibration data appropriate to the epoch of observation
  (primarily \ciao 4.6 and the corresponding calibration database,
  version 4.5.9, though some recent reprocessing was done with \ciao
  4.10 and CALDB 4.7.8).\footnote{Some recent calibration updates have
    retroactive changes in the effective area of order 5\% at
    $12\mang$ due to revisions in the contamination model.  There are
    no other significant differences in recent (post observation
    epoch) \ciao versions which affect event processing, spectral
    extraction, or line characteristics. For these data and our
    purposes, the effective area revisions are not important, so we
    have not re-extracted and re-fit data with the most recent
    calibrations.}}
The counts spectra are composed of four orders per source per
observation: the $\pm1$ orders for each grating type, the \meg and
\heg, which have different efficiencies and resolving powers.  The
default binning over-samples the instrumental resolution by about a
factor of four.

Observation-specific calibration files are required for analysis to
convolve a model flux spectrum with the instrumental response to
produce model counts.  \ciao programs were used to make the effective
areas (``Auxiliary Response File'', or \arf) and the spectral
redistribution and extraction-aperture efficiency files (``Response
Matrix File'', or \rmf) for each spectral order for each source
\citep[see][for a detailed definition of the response]{DavisJE:2001b}.

\begin{figure}[!htb]
  \centering\leavevmode
 \includegraphics[width=0.98\columnwidth, viewport = 30 180 520 550]{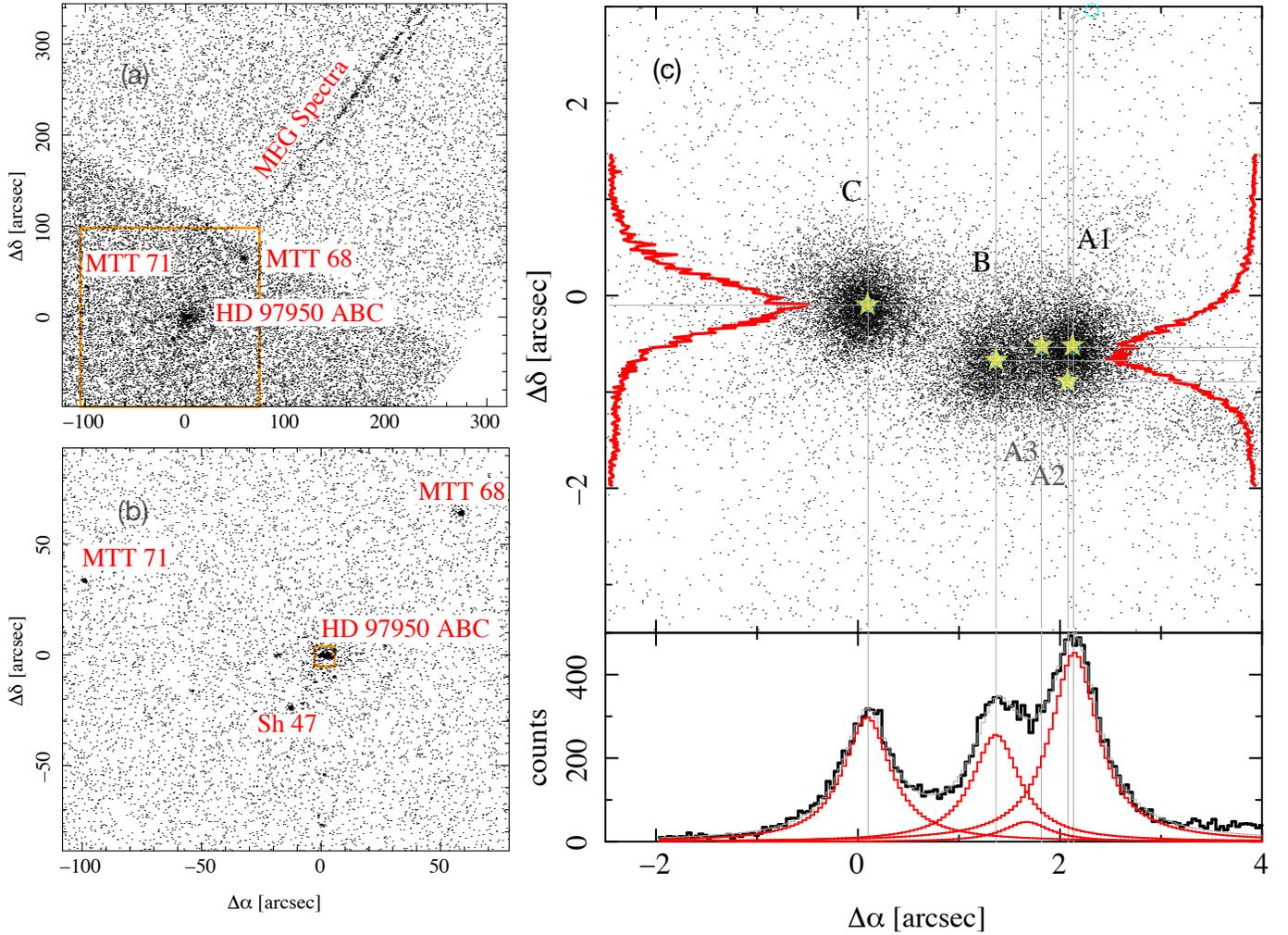}
  \caption{Images of the field for decreasing scales.  The top left
    panel is a large-scale view with the cluster core in the lower
    left and the MEG spectrum running diagonally to the upper right.
    The lower left is an expanded scale.  The left panels are from the
    HETG observation (observation ID 13266).  The right panel shows
    the deeper ACIS-I data (ObsIDs 12328, 12329), which better shows
    the event distribution of the HD~97950 core. Known star positions
    are marked, and marginal histograms along the axes show the X-ray
    event distributions ($y$-scales for the upper-panel marginal
    histograms are arbitrary).  Positions of the components are from
    \citet{Drissen:Moffat:al:1995}. For the $y$-axis histograms, we
    have separated events for C from A1-B.  In all panels, east is
    left and north is up.}
  \label{fig:fieldzooms}
\end{figure}

To extract spectra of multiple sources, we need to customize source
positions and extraction regions to minimize confusion. In
Figure~\ref{fig:fieldzooms} we show X-ray images of the field at
successively decreasing scales.  In the rightmost panel, we overlay
positions of the optical components from
\citet{Drissen:Moffat:al:1995}.   Properties of objects relevant to
this study are given in Table~\ref{tbl:srcprops}.
%
%
\begin{deluxetable}{ccccccccc}
  \tablecaption{Properties of Selected NGC 3603 Members}\label{tbl:srcprops}
  \tablehead{
    \colhead{Star}&
    \colhead{Spectral}&
    \colhead{Binarity, }&
    \colhead{Mass}&
    \colhead{$L_\mathrm{bol}\tablenotemark{a}$}&
    \colhead{$v_\infty$}&
    \colhead{$\gamma$}&
    \colhead{$K$}&
    \colhead{$f_\mathrm{x}\tablenotemark{b}$}\\
    \colhead{}&
    \colhead{Type}&
    \colhead{Period (d)}&
    \colhead{($M_\odot$)}&
    \colhead{($10^6\,L_\odot$)}&
    \colhead{($\kms$)}&
    \colhead{($\kms$)}&
    \colhead{($\kms$)}&
    \colhead{($\eflux$)}
  }
  \startdata
  A1& 
  $\mathrm{WN6h+WN6h}$&
  SB2, 3.77&
  116, 89&
  4.0&
  2700&
  153&
  330, 433&
  $5.74\times10^{-13}$\\
  A2&
  O3 V &
  -&
  -&
  1.2&
  -&
  -&
  -&
  -\\
  A3&
  O3 III (f*)&
  -&
  -&
  0.8&
  -&
  -&
  -&
  -\\
  B&
  WN6h&
  single&
  89&
  2.9&
  2700&
  167&
  -&
  $2.69\times10^{-13}$\\
  C&
  $\mathrm{WN6h+?}$&   
  SB1, 8.9&
  62&
  2.2&
  2500&
  186&
  200&
  $1.08\times10^{-12}$\\
  MTT~68&
  O2 If*&
  vis., $0.38''$&
  $150$&
  0.4&
  -&
  -&
  -&
  $1.28\times10^{-12}$\\
  MTT~71&
  O4 III&
  -&
  $80$&
  1.5&
  -&
  -&
  -&
  $2.46\times10^{-13}$\\
  Sher~47&
  O4 IV (f)&
  -&
  -&
  0.5&
  -&
  -&
  -&
  $3.53\times10^{-13}$\\
  \enddata
  \tablecomments{Designations A-B-C refer to the sub-components of
    HD~97950. \newline
    Data sources:
    \citet{
      Drissen:Moffat:al:1995,
      Crowther:Dessart:1998,
      vanderhucht:2001,
      Melena:Massey:al:2008,
      Schnurr:Casoli:al:2008,
      Crowther:al:2010,
      Roman-Lopes:2013b,
      Roman-Lopes:al:2016,
      Apellaniz:al:2016}.
    \tablenotetext{a}{Luminosities for A1, B, and C are from
    \citet{Crowther:al:2010}; others are from
    \citet{Crowther:Dessart:1998} but re-scaled to a distance of
    $7.6\kpc$ (from their value of $10.1\kpc$).}
    \tablenotetext{b}{Flux is the total band ($0.5$--$8.0\kev$) from
      \citet{Townsley:al:2014}, with a caveat that these were
      significantly saturated by CCD event pile-up.}
  }
\end{deluxetable}

From the marginal histograms, we can see that the X-rays in the core
are dominated by components A1, B, and C, and that in this
observation, C is fainter than A1.  The dispersion direction is
roughly diagonal, and the MEG trace of A1+B can be seen faintly in
the upper-left panel of Figure~\ref{fig:fieldzooms} extending to the
upper right, along with the bright sources MTT~68 at about
$1.4\,\mathrm{arcmin}$ northwest of the core \citep{Melnick:al:1989,
  Roman-Lopes:2013b}, and MTT~71 at about $1.7\,\mathrm{arcmin}$
northeast of the core. The \hetg resolution is not affected by
off-axis angles below $2\arcmin$.

\begin{figure}[!htb]
  \centering\leavevmode
 \includegraphics*[width=0.50\columnwidth, viewport = 0 0 550 520]{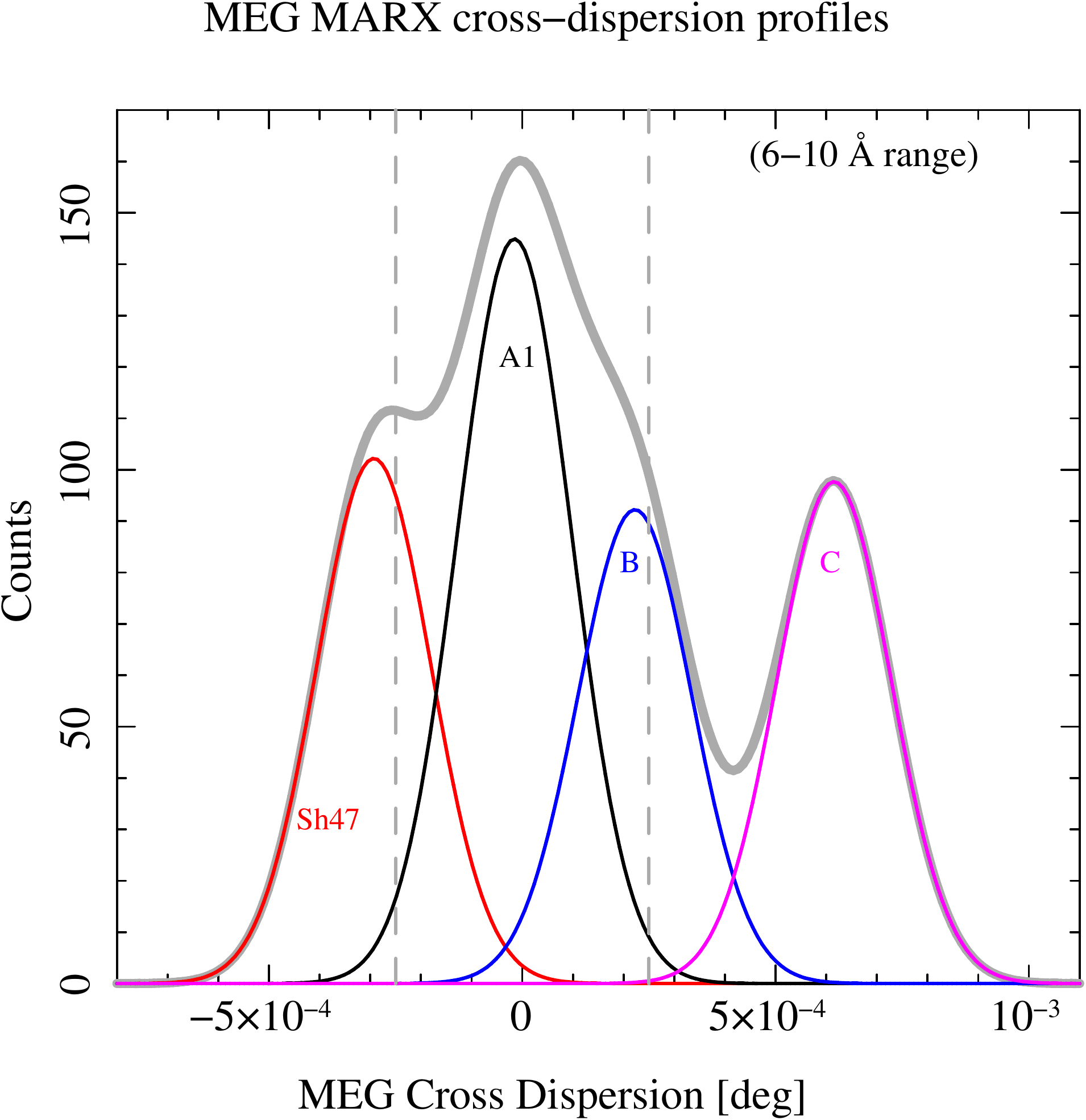}
 \caption{
   { Cross-dispersion profiles for the A1+B extraction in the
     $6$--$10\mang$ range of the MEG order, from MARX simulations.
     Vertical dashed lines show the limits of the extraction region
     for A1, which includes tails of Sh~47 and B.}
   }
  \label{fig:xdisp}
\end{figure}

We used MARX\footnote{\url{http://space.mit.edu/cxc/marx}}
\citep{Davis:al:MARX:2012} to simulate the field and to assess source
confusion in detail.  While components A1 and B are marginally
resolved in zeroth order, they are not in the dispersed spectrum, due
to the additional grating astigmatic profile in the cross-dispersion
direction.  {The wavelength offset between A1 and B, projected
  along the MEG direction, is $1.46\,$pixels ($0.016\mang$), or about
  $800\kms$ at $6\mang$.  With the extraction centered on A1,
  combination of positive and negative orders would in principle
  result in B's contributions being offset by $\pm800 \kms$ then
  summed with A1.  However, A1 is twice as bright as B, and using a
  narrow extraction region reduces B's contribution further, as does
  including the well separated HEG spectra. Experiments in fitting
  faked data with offsets showed that line centroids are unaffected
  (as expected), but the line shapes become a bit flatter. Even with
  some broadening due to confusion, we will see in
  Section~\ref{sec:modeling} that the line widths are still much
  larger than we expect from source confusion.

  In Figure~\ref{fig:xdisp}, we show cross-dispersion profiles as fit
  to MARX simulations, using the $6$--$10\mang$ region of the MEG
  spectrum of the A1+B extraction.  The counts in the A1 extraction
  region in the MEG arm alone is about 60\% from A1.  }

\begin{figure}[!htb]
  \centering\leavevmode
 \includegraphics*[width=0.85\columnwidth, viewport = 0 0 550 325]{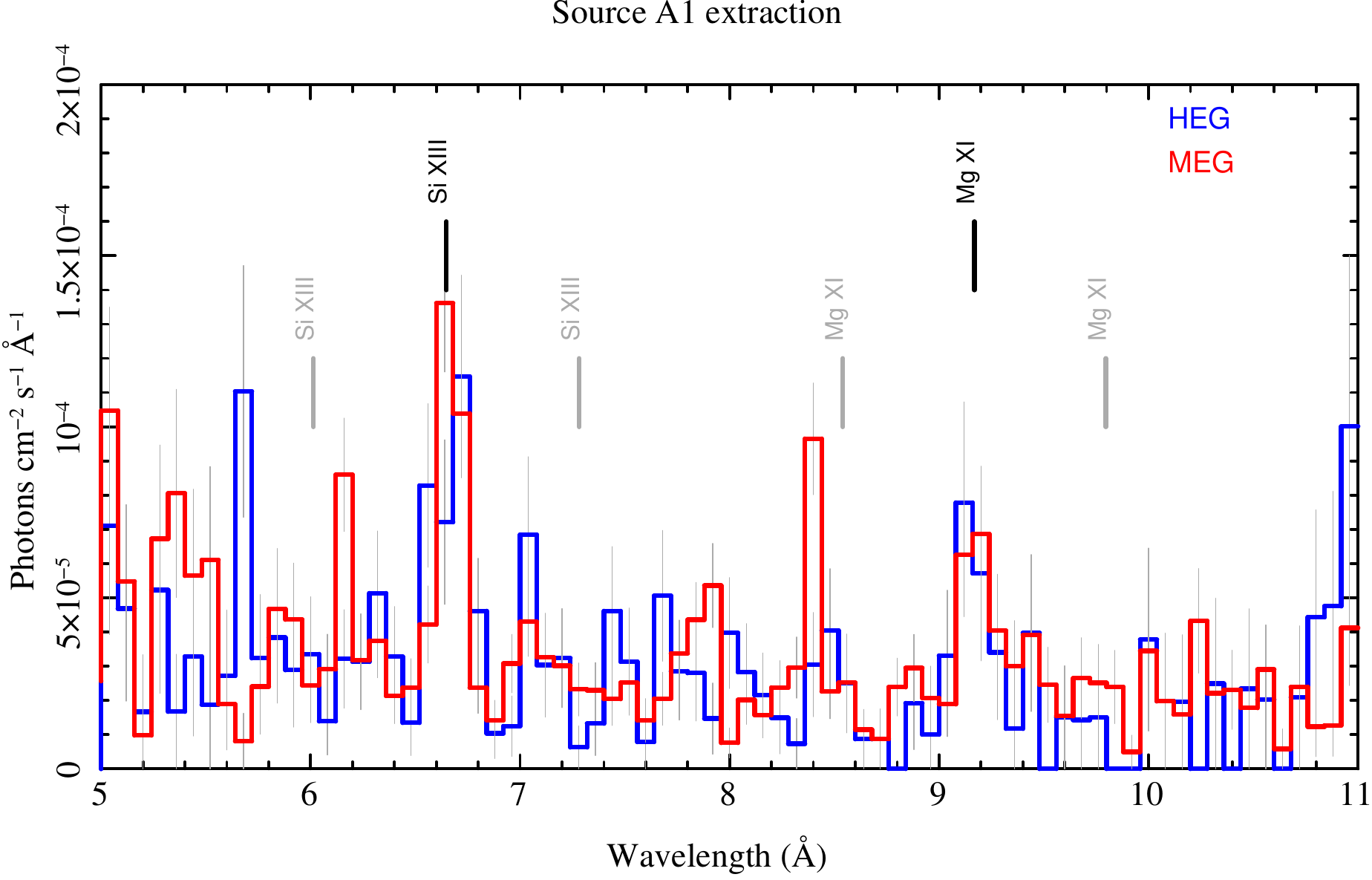}
 \caption{ { The A1+B flux spectra for HEG (blue) and
     MEG (red), each of which have positive and negative orders
     combined.  If Sh~47 significantly contaminated the MEG spectrum
     of A1+B, the HEG and MEG fluxes would not agree.  We also see no
     significant emission features at the expected positions for the
     Sh~47 spatial offset; we mark the strongest expected lines of
     \eli{Si}{13} and \eli{Mg}{12} for the A1+B extraction scale in
     black, and to either side in gray, where Sh~47 confusing lines
     would be.  Spectra have been binned to $0.08\mang/\mathrm{bin}$.}
 } \label{fig:shconf}
\end{figure}

We also analyzed MTT~68, MTT~71, and Sher~47.  The latter, however, is
aligned with the A1+B MEG spectrum, so its MEG spectrum is not
useful.  
{In turn, its MEG counts contaminate the A1+B spectrum, though it is
  offset enough, by $0.63\mang$, that any emission lines will not
  overlap those of A1+B. It is somewhat fainter---we estimate that
  Sh~47 could contribute about 12\% of the A1+B HEG plus MEG counts
  spectrum.  As a further test of confusion from Sh 47, we can compare
  the HEG and MEG flux of A1+B, which should differ if Sh~47
  contaminates the MEG arm, and we can look for offset spectral
  features. In Figure~\ref{fig:shconf}, we show the HEG and MEG
  fluxes\footnote{Flux spectra are derived by dividing by the counts
    expected for a flat spectrum, that is, by the integral over the
    response. This is an approximation to the flux, but still includes
    the instrumental broadening.  This was done in the ISIS analysis
    package, and full details are given in the manual (see \url{
<http://space.mit.edu/cxc/software/isis/manual.pdf\#page.76>}).} for the A1+B extraction. The fluxes
  agree, and there are no features at the expected offsets.  Hence, we
  conclude that the limited extraction region widths and order-sorting
  have mitigated the contributions from Sh~47.}

\begin{figure}[!htb]
  \centering\leavevmode
  \includegraphics[width=0.45\columnwidth]{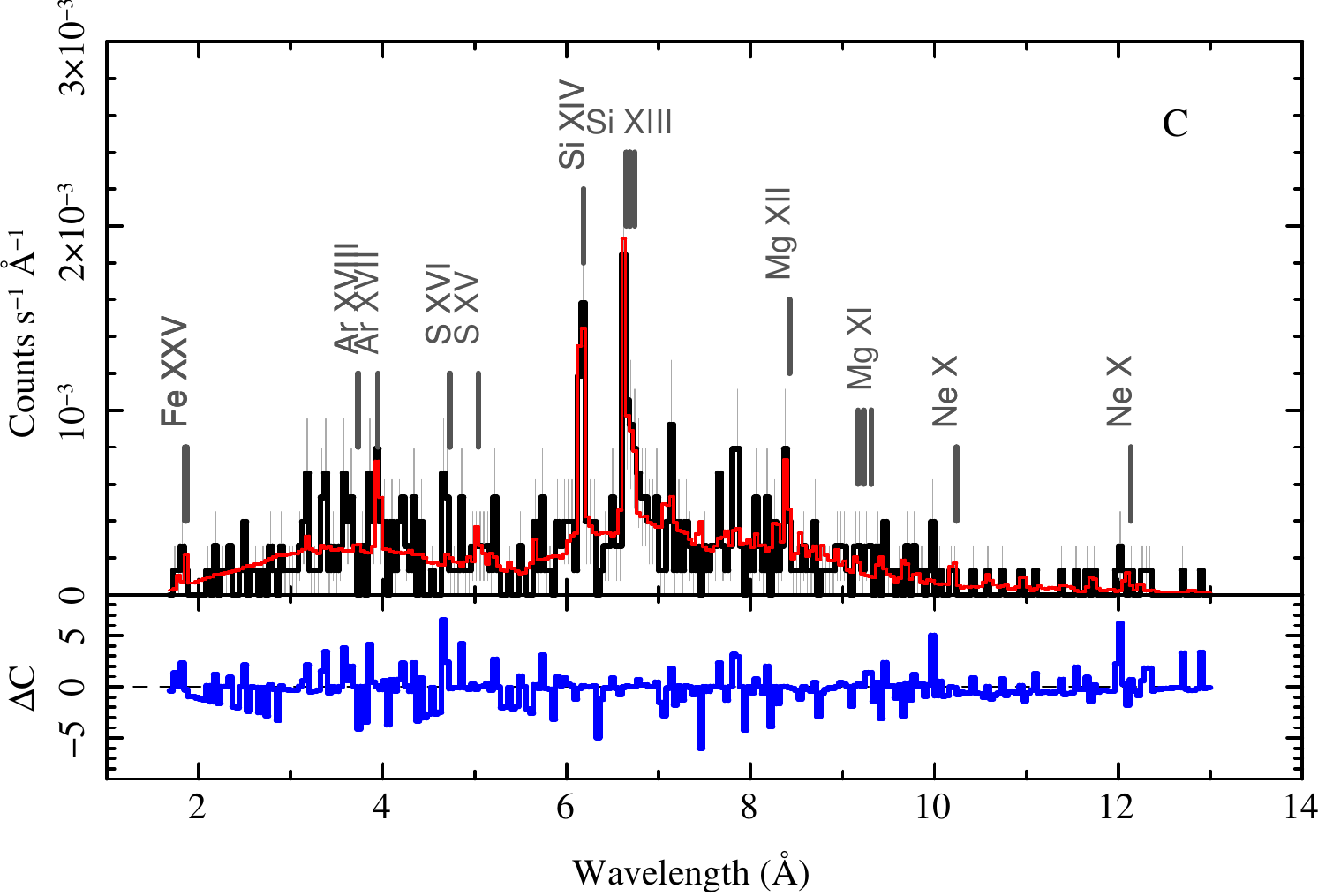}
  \includegraphics[width=0.45\columnwidth]{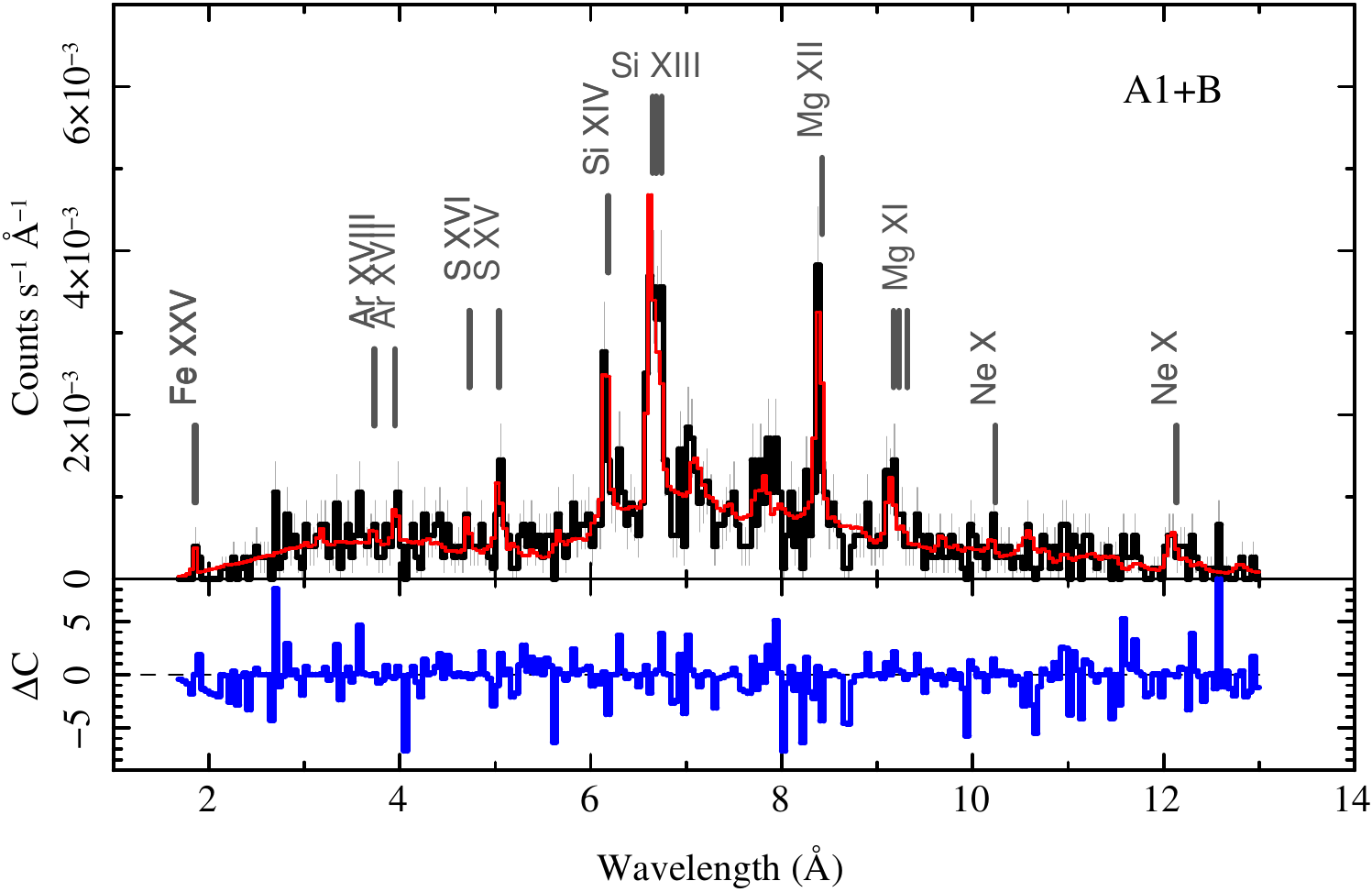}
  \includegraphics[width=0.45\columnwidth]{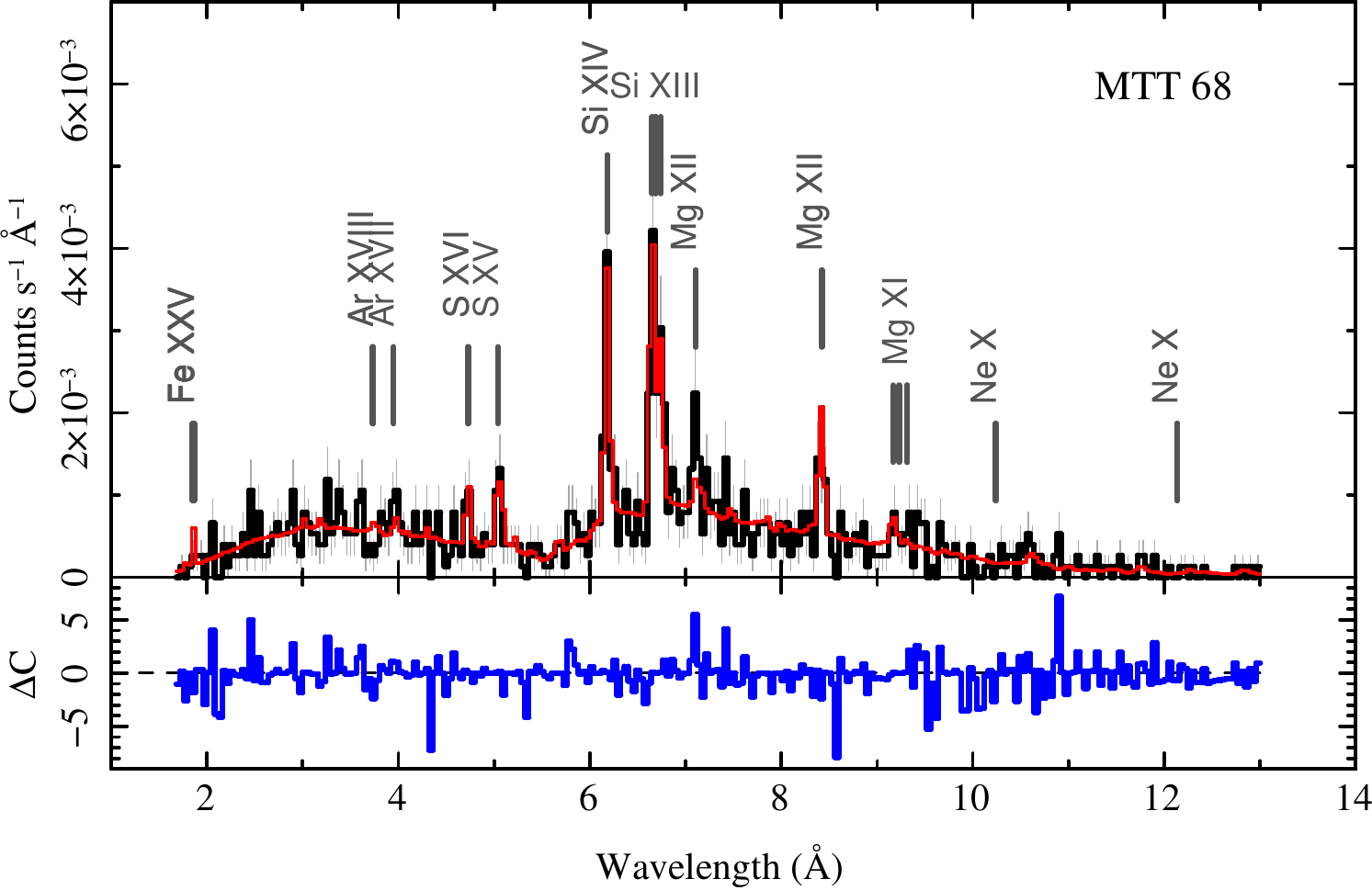}
  \includegraphics[width=0.45\columnwidth]{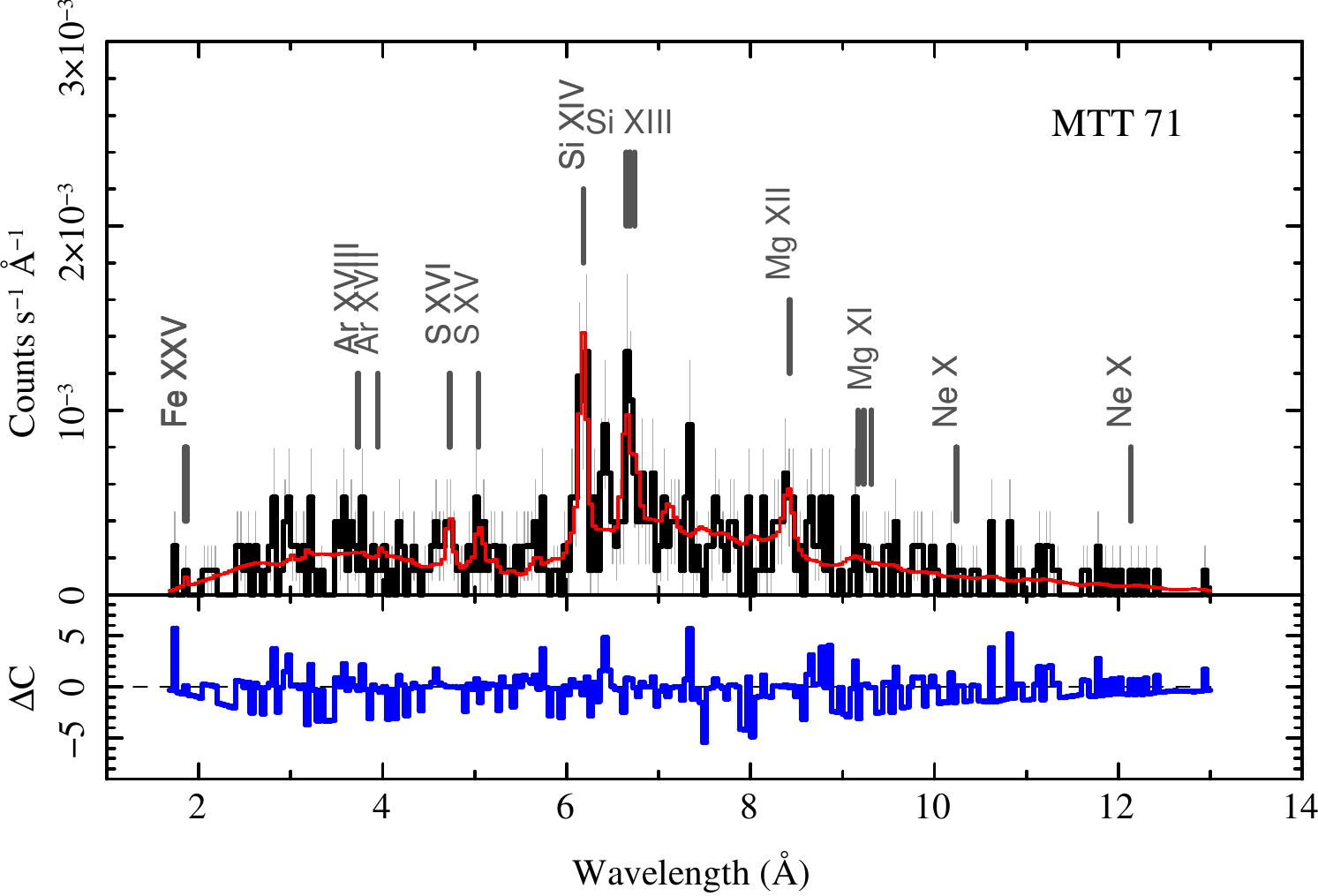}
  \caption{Summary \hetgs counts spectra: combined \heg and \meg first
    orders.  The locations of some spectral features, primarily H- and
    He-like ions, are marked.  The red curve is a two-temperature
    AtomDB \citep{Foster:Smith:Brickhouse:2012}
    model from a global fit, whose scaled residuals (as the
    $C$-statistic) are shown in the lower portion of each panel.}
  \label{fig:summaryspec}
\end{figure}

In Figure~\ref{fig:summaryspec}, we show the combined flux spectra
for the \heg and \meg first orders of components A1+B, C, 
MTT~68, and MTT~71.  Count rates and fluxes (which are largely
model-independent) are given in Table~\ref{tbl:rates}.

{
In this relatively short, single pointing, we cannot totally mitigate
source confusion.  In the future, we hope to obtain deeper exposures
at multiple roll angles which will allow us to better extract
unconfused spectra.
}

\begin{deluxetable}{crrr}
  \tablecaption{Source Rates for Selected NGC 3603 Members}
  \tablehead{
    \colhead{Object}&
    \colhead{$r_0$}&
    \colhead{$r_{\pm1}$}&
    \colhead{$f_\mathrm{x}$}
  }
  \startdata
  A1& 8.84 & -- & -- \\
  B&  4.72 & 26.0\tablenotemark{a} &      1.31\tablenotemark{a}\\
  C&  6.81 & 9.47 &     0.66\\
  MTT~68&  15.33 &     22.0 &      1.37\\
  MTT~71& 4.58 &     9.32 &      0.51\\
  Sher~47\tablenotemark{b}&  7.49 & 3.65 & 0.63\\
  \enddata
  
  \tablenotetext{a}{\hetgs dispersed spectral rates are for A1+B.}
  \tablenotetext{b}{First orders and flux are from HEG only.}
  
  \tablecomments{
    Rates for zeroth order ($r_0$) and $\pm1^\mathrm{st}$ orders
    ($r_{\pm1}$) are given in $\mathrm{cts\,ks\mone}$.\newline
    Flux in the dispersed spectra, $f_x$, is in $10^{-12}\eflux$, using
    the model evaluated over $1$--$40\mang$.
  }\label{tbl:rates}
\end{deluxetable}

{
  We have examined the background rate by extracting the relatively
  isolated MTT~68 with standard width spatial masks, which provide 10
  times the width of the source region for spectrally-adjacent
  background regions.  The rate is less than about
  $10^{-4}\,\cts\mang^{-1}$ for standard extraction regions, and is
  negligible for the grating spectral analysis.
}

{ We have not included zeroth orders in our analysis.  While
  the HETG zeroth order effective area exceeds that of the combined
  first orders above $2\kev$ (below $6\mang$) by up to a factor of 3
  (with a median factor of 1.6), there are uncertainties in the
  calibration of order 5--10\% between zeroth and first orders.
  Furthermore, we would require small extraction regions for the close
  sources, C, B, and A1 and the aperture correction introduces another
  potentially larger systematic uncertainty.  As can be seen from
  inspection of the counts below $6\mang$ in
  Figure~\ref{fig:summaryspec}, we do have enough signal in the
  dispersed spectrum for characterization of the highest temperature
  plasmas.  Since we will be primarily interested in the resolved
  emission lines, we will not further consider the zeroth order in our
  analysis.  }

\section{Spectral Modeling}\label{sec:modeling}

To characterize the spectra, we fit two temperature, foreground
absorbed {AtomDB \citep{Foster:Smith:Brickhouse:2012}
plasma emissivity models.}\footnote{The atomic database, AtomDB (\url{www.atomdb.org}),
  is produced by the {\it Astrophysical Plasma Emission Code}, or APEC,
  which is also commonly referred to as the {\it Astrophysical Plasma
  Emission Database}, or APED.  These are often used interchangeably
  when referring to spectral emissivity models derived from APEC.}
This is not meant to imply that this is an appropriate physical model
for the X-ray emission from a stellar wind.  A wind is accelerating
and possibly stratified: emission can originate from regions of
different velocities, temperatures, and overlying continuum absorption
optical depths.  Winds can also be clumped.  These all affect the
X-ray energy distribution and line profiles \citep{Owocki:Cohen:2001,
  Oskinova:Feldmeier:Hamann:2004, Cohen:Wollman:al:2014}.  Instead,
our intent is to provide a simple characterization of the emitting
plasma appropriate to the quality of the data, and to measure emission
line parameters which are independent of global plasma models.

Because of the heavy foreground absorption of about $10^{22}\cmmtwo$
\citep{Townsley:al:2014, Romano:al:2008, Moffat:Corcoran:al:2002}, we
are not sensitive to cooler plasma temperatures of a few MK.  Since we
see emission lines of Si and Mg, we have temperatures near $10\mk$,
and the continuum present between $2$--$6\mang$ indicates temperatures
in the tens of $\mk$, hence a $2T$ model should suffice for general
characterization of the plasma model.  In addition
to foreground absorption, there could be intrinsic absorption within
the stellar wind itself, hence, we let the column be fit.  

Since the emission lines are of primary interest, we also fit a global
``turbulent'' broadening parameter (i.e., constant velocity width),
and a Doppler shift.  These allow us to characterize the line width
and apparent offset of the centroid, which might be due to skewness of
the profile \citep[see for example][]{Owocki:Cohen:2001}.  

Finally, we also allowed elemental abundances for Fe, Ar, S, Si, Mg,
and Ne to be free.  While these parameters are related to the physical
abundances, they also compensate for systematic errors in plasma
temperatures as per limitations of the adopted two-component model.
We modified the \citet{Asplund:Grevesse:al:2009} appropriately for the
depleted H fraction of WNh stars given by
\citet{Crowther:Dessart:1998} {in which the H to He number
  ratio is 6, or by mass fraction, 0.59 H.\footnote{ In the usual
    by-number decimal logarithmic scale in which the hydrogen
    abundance is 12, the WNh abundances of the elements relevant here
    are $\mathrm{He} = 11.22$, $\mathrm{Ne} = 8.02$,
    $\mathrm{Mg} = 7.69$, $\mathrm{Si} = 7.60$, $\mathrm{S} = 7.21$,
    $\mathrm{Ar} = 6.49$, and $\mathrm{Fe} = 7.59$.}  Only the
  relative abundances of Mg, Si, and Fe are reasonably constrained,
  since they have the strongest or most numerous (in the case of Fe)
  lines.}

We fit the $1.7$--$13\mang$ region using a Powell minimization method
and a Cash (maximum likelihood) statistic.  These global fits and
residuals are shown in Figure~\ref{fig:summaryspec}.  Model parameters
are given in Tables~\ref{tbl:specpar} and \ref{tbl:specparabund}.
Fluxes of the models, since they characterize the empirical shape of
the observed spectrum (and are largely model independent), are given
in Table~\ref{tbl:rates}.  Luminosities, which depend critically on
assumed absorption (here all assumed in the foreground) are given in
Table~\ref{tbl:specpar}, using a distance of $7.6\kpc$
\citep{Melena:Massey:al:2008}.

%
\begin{deluxetable}{c|ccc|ccc|ccc|ccc|ccc|cc} 
  \tablecaption{Spectral Model Parameters}
  \tabletypesize{\scriptsize}
  \tablehead{
    \colhead{Object}&
    \colhead{$N_H$}&
    \colhead{$\sigma_\mathrm{lo}$}&
    \colhead{$\sigma_\mathrm{hi}$}&
    \colhead{$T_1$}&
    \colhead{$\sigma_\mathrm{lo}$}&
    \colhead{$\sigma_\mathrm{hi}$}&
    \colhead{$T_2$}&
    \colhead{$\sigma_\mathrm{lo}$}&
    \colhead{$\sigma_\mathrm{hi}$}&
    \colhead{$EM_1$}&
    \colhead{$\sigma_\mathrm{lo}$}&
    \colhead{$\sigma_\mathrm{hi}$}&
    \colhead{$EM_2$}&
    \colhead{$\sigma_\mathrm{lo}$}&
    \colhead{$\sigma_\mathrm{hi}$}&
    \colhead{$L_x$\tablenotemark{a}}&
    \colhead{$\log(L_x/L_\mathrm{bol})$}\\
    \multicolumn{1}{c}{\ }&
    \multicolumn{3}{c}{($10^{22}\cmmtwo$)}&
    \multicolumn{3}{c}{(MK)}&
    \multicolumn{3}{c}{(MK)}&
    \multicolumn{3}{c}{($10^{56}\cmmthree)$}&
    \multicolumn{3}{c}{($10^{56}\cmmthree)$}&
    \multicolumn{1}{c}{($10^{34}\lum$)}&
    \multicolumn{1}{c}{(dex)}
  }
  \startdata
  A1+B&     1.3&   0.9&    1.9&   8.7&  6.4&   9.5& 36.5&  27.9& 48.2& 8.0& 4.9& 13.2& 4.4&  3.5& 5.6&  3.60&-5.8\\
  C&	      2.6&   1.9&    3.0:&  12.1& 10.7& 13.4& 92.1&  40.5& 100:& 5.3& 2.7& 9.0&  1.2&  0.8& 2.1&  2.59&-5.5\\
  MTT~68&     1.5&   1.2&    2.1&  10.8&  9.0&  12.6& 60.7&  43.5& 100:& 2.9& 1.7& 7.0&  4.3&  3.3& 5.3&  2.17&-4.8\\
  MTT~71&     0.8&   0.8&    1.5&  12.6&  7.3&  18.6& 50.2&  33.0& 100:& 0.7& 0.2& 3.0&  1.8&  0.9& 2.4&  0.60&-6.0\\
  Sher~47\tablenotemark{b}&     1.0:&  --- &  ---  &  6.9 & 2.8 & 11.1 & 27.6&  22.1&  35.8& 1.2& 0.3& 5.9&  3.2&  2.5& 4.0&  1.20:&-5.2:\\
  \enddata              
  \tablenotetext{a}{Assumes a distance of $7.6\kpc$.}
  \tablenotetext{b}{Only the HEG spectrum was fit, and the value of
    $N_\mathrm{H}$ was assumed.}
  \tablecomments{Columns labeled $\sigma_{lo}$, $\sigma_{hi}$ are the
    90\% confidence limits for the preceding parameter.  A colon
    (``:'') indicates an uncertain value, as in an error limit which
    did not converge.  The model fit was of the form 
        $\mathrm{ ( AtomDB(1) + AtomDB(2) ) * PHABS(1)}$.} \label{tbl:specpar}
\end{deluxetable}
%

\begin{deluxetable}{c|ccc|ccc|ccc|ccc|ccc|ccc} %
  \tablecaption{ {Spectral Model Parameters: Relative Abundances}}
  \tabletypesize{\small}
  \tablehead{
    \colhead{Object}&
    \colhead{Ne}&
    \colhead{$\sigma_\mathrm{lo}$}&
    \colhead{$\sigma_\mathrm{hi}$}&
    \colhead{Mg}&
    \colhead{$\sigma_\mathrm{lo}$}&
    \colhead{$\sigma_\mathrm{hi}$}&
    \colhead{Si}&
    \colhead{$\sigma_\mathrm{lo}$}&
    \colhead{$\sigma_\mathrm{hi}$}&
    \colhead{S}&
    \colhead{$\sigma_\mathrm{lo}$}&
    \colhead{$\sigma_\mathrm{hi}$}&
    \colhead{Ar}&
    \colhead{$\sigma_\mathrm{lo}$}&
    \colhead{$\sigma_\mathrm{hi}$}&
    \colhead{Fe}&
    \colhead{$\sigma_\mathrm{lo}$}&
    \colhead{$\sigma_\mathrm{hi}$}
  }
  \startdata
%
%
   AB& 1.7& 0.6& 3.9& 0.9& 0.6& 1.4& 1.1& 0.7& 1.7& 1.5& 0.7& 2.8& 2.8& 0.1& 6.6& 0.9& 0.4& 1.9\\
    C& 6.6& 0.8& 6.6& 0.6& 0.2& 1.3& 1.1& 0.6& 1.8& 0.4& 0.1& 1.5& 5.1& 1.3& 6.6& 1.7& 0.7& 3.3\\
MTT68& 0.1& 0.1& 3.8& 1.3& 0.7& 2.2& 2.5& 1.3& 3.3& 3.3& 1.4& 3.3& 1.5& 0.1& 6.6& 0.9& 0.3& 1.6\\
MTT71& 0.3& 0.1& 5.0& 1.0& 0.2& 2.6& 2.5& 0.9& 3.3& 3.1& 0.6& 3.3& 0.1& 0.1& 6.6& 0.3& 0.1& 1.0\\
  \enddata              
  \tablecomments{Abundances are by number fraction relative to the WNh
    values of \citet{Crowther:Dessart:1998}.  Columns labeled
    $\sigma_{lo}$, $\sigma_{hi}$ are the 90\% confidence limits for
    the preceding parameter.  Only Mg, Si, and Fe are resonably
    constrained.  The others are listed only because they are formal
    values of the fit.  Sh~47 is not included because it had too weak
    a spectrum to fit abundances, which were assumed to be 1.0.
  } \label{tbl:specparabund}
\end{deluxetable}
%

In addition to using Gaussian profiles, we used a simple analytic
profile characteristic of asymptotic flow of thick wind
\citep{Ignace:2001} as successfully applied to the \hetgs spectrum of
the Wolf-Rayet star, WR~6 \citep{Huenemoerder:al:2015}.  This profile
has two shape parameters, the terminal velocity ($v_\infty$), and a
shape, $q>-1$; $q=0$ is the nominal wind expansion model which creates
a ``shark-fin'' shaped profile; $q=-1$ results in a flat-topped
profile (as would be emitted by a physically thin expanding shell; and
$q>0$ becomes increasingly steep, rapidly falling from the sharp blue
wing at $-v_\infty$ (see \citet{Huenemoerder:al:2015} for example
profiles and the analytic functional form).  This line model was {\em
  not} allowed to have a free Doppler shift; line centers were instead
frozen at their relatively small line-of-sight velocities.

{ All fitting and modeling was done using the {\it Interactive
    Spectral Interpretation System}
  (ISIS\footnote{\url{http://space.mit.edu/cxc/software/isis}};
  \cite{Houck:00}), which provides interfaces to AtomDB and Xspec
  models.  }

\begin{figure}[!htb]
  \centering\leavevmode
  \includegraphics[width=0.45\columnwidth, viewport=0 0 340 320]{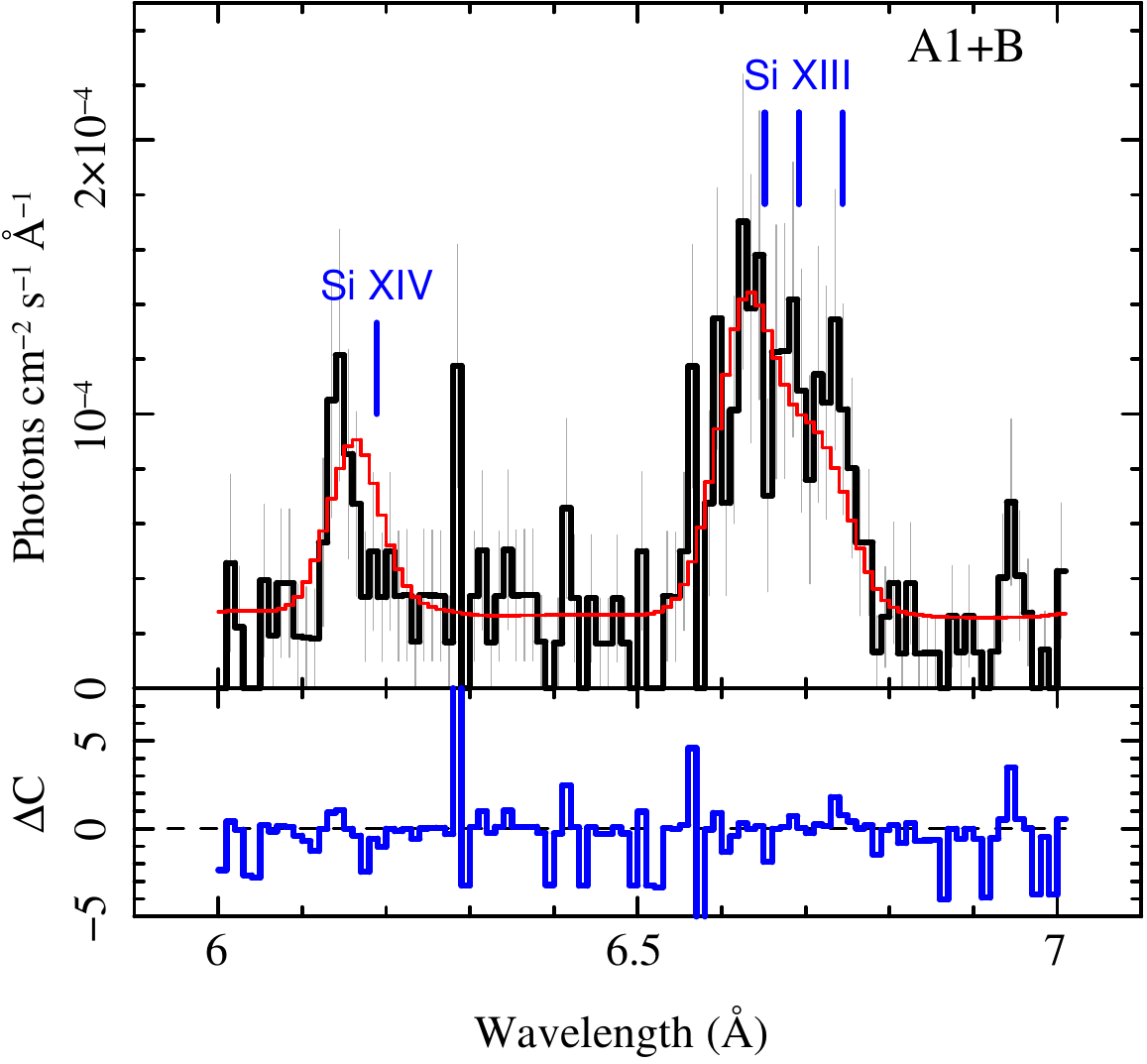}
  \includegraphics[width=0.45\columnwidth, viewport=0 0 340 320]{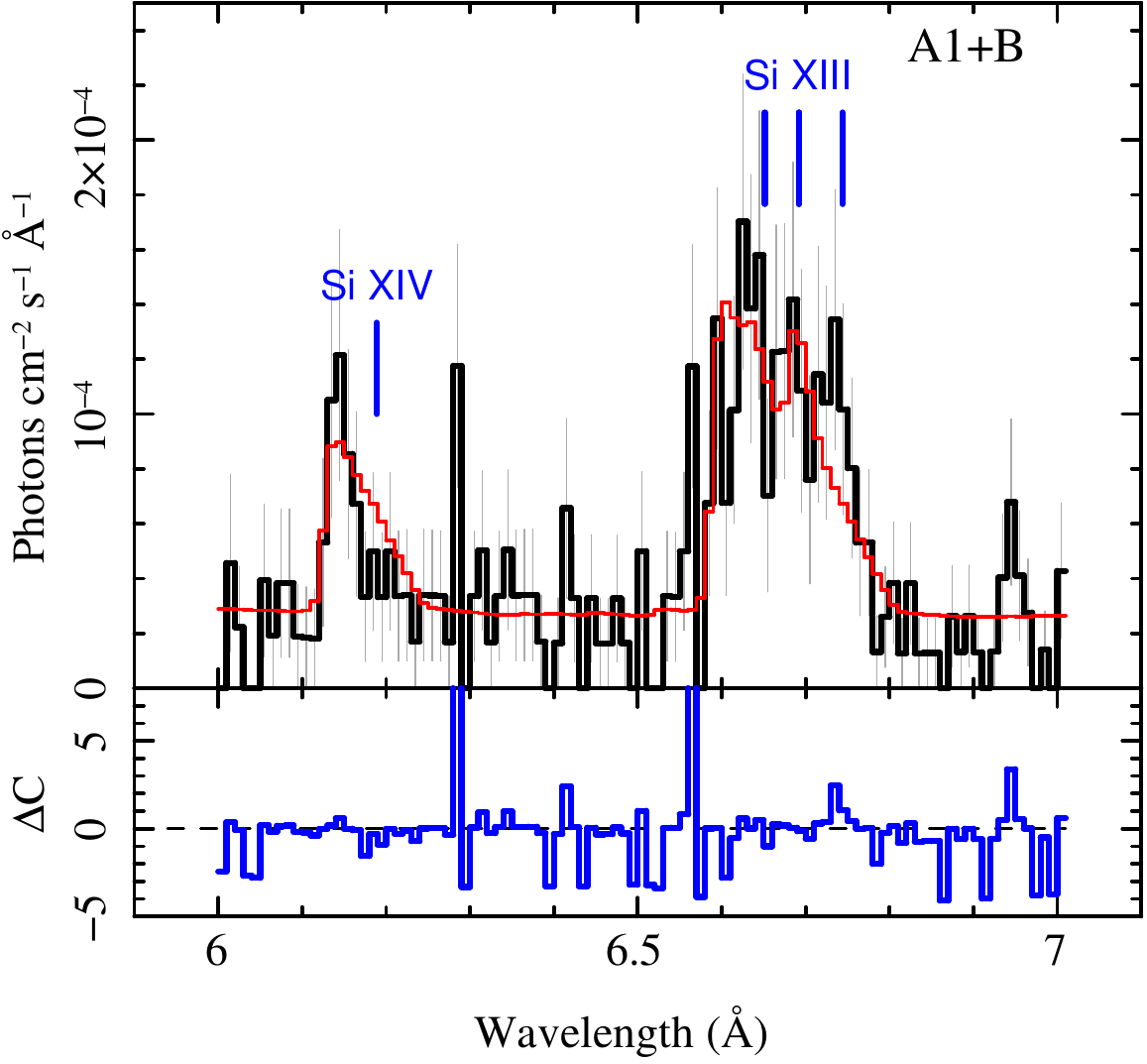}
  \caption{Gaussian (left) and wind (right) profile AtomDB plasma fits to the to
    A1+B spectra.  Black are the observed counts for the combined \heg
    and \meg first orders.  The red curve is the AtomDB model with
    Doppler shifted, Gaussian-broadened (left), or an asymptotic wind
    (right) profile.  The \eli{Si}{14} H-like ($6.18\mang$) and
    \eli{Si}{13} He-like lines ($6.65$, $6.69$, and $6.74\mang$) rest
    wavelengths are marked.  The Si lines are broader than instrumental, and
    also have a blueshifted centroid.}
  \label{fig:sifits}
\end{figure}

In Figure~\ref{fig:sifits} we show the \eli{Si}{14} to \eli{Si}{13}
region twice, once fit with the AtomDB models using Gaussian profiles,
and again as fit with the asymptotic thick wind profiles.  The
wind-profile is convenient because it clearly delimits the maximum
extent of the blue wing, but we cannot claim that it is better than a
Gaussian for these stars.  Our general qualitative impression is that
it seems to better match \eli{Si}{14} than a Gaussian.

\begin{figure}[!htb]
  \centering\leavevmode
  \includegraphics*[width=0.45\columnwidth, viewport=0 0 435 425]{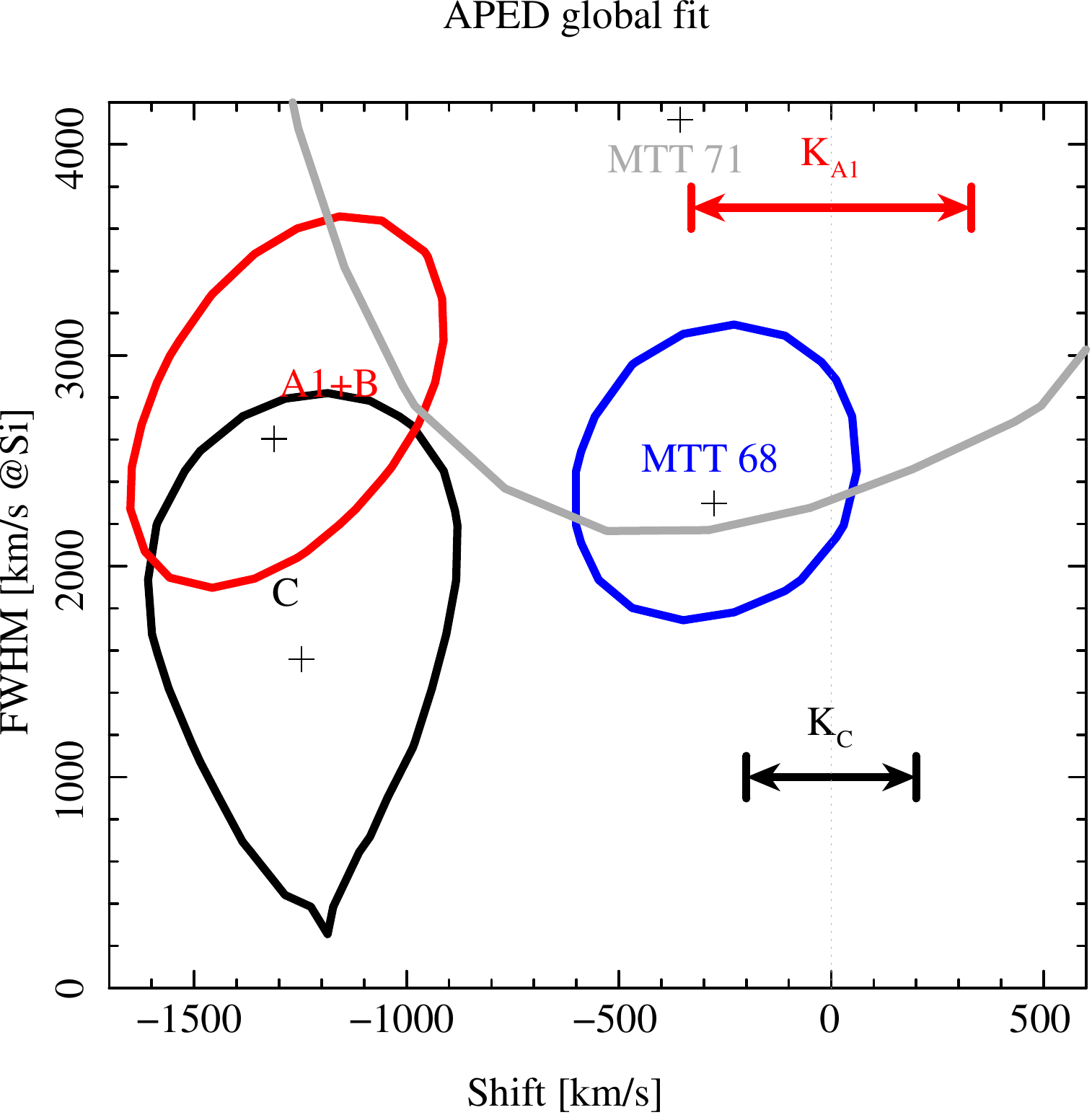}
  \caption{Gaussian-broadened, AtomDB fits to the $1.7$--$13\mang$
    region, confidence contours of centroid vs line width ($90\%$ 
    contours). $K_x$ denotes the expected velocity amplitude of the
    binaries, A1 and C.  Since the Doppler shifts are much larger than
    radial velocity amplitudes, they are due to line asymmetry, not
    orbital motion. MTT~68 is unshifted, and so is likely due to a
    thinner wind.  MTT~71 is poorly constrained, but does have
    broadened lines.}
  \label{fig:confmapa}
\end{figure}

\begin{figure}[!htb]
  \centering\leavevmode
  \includegraphics*[width=0.45\columnwidth, viewport=0 0 435 430]{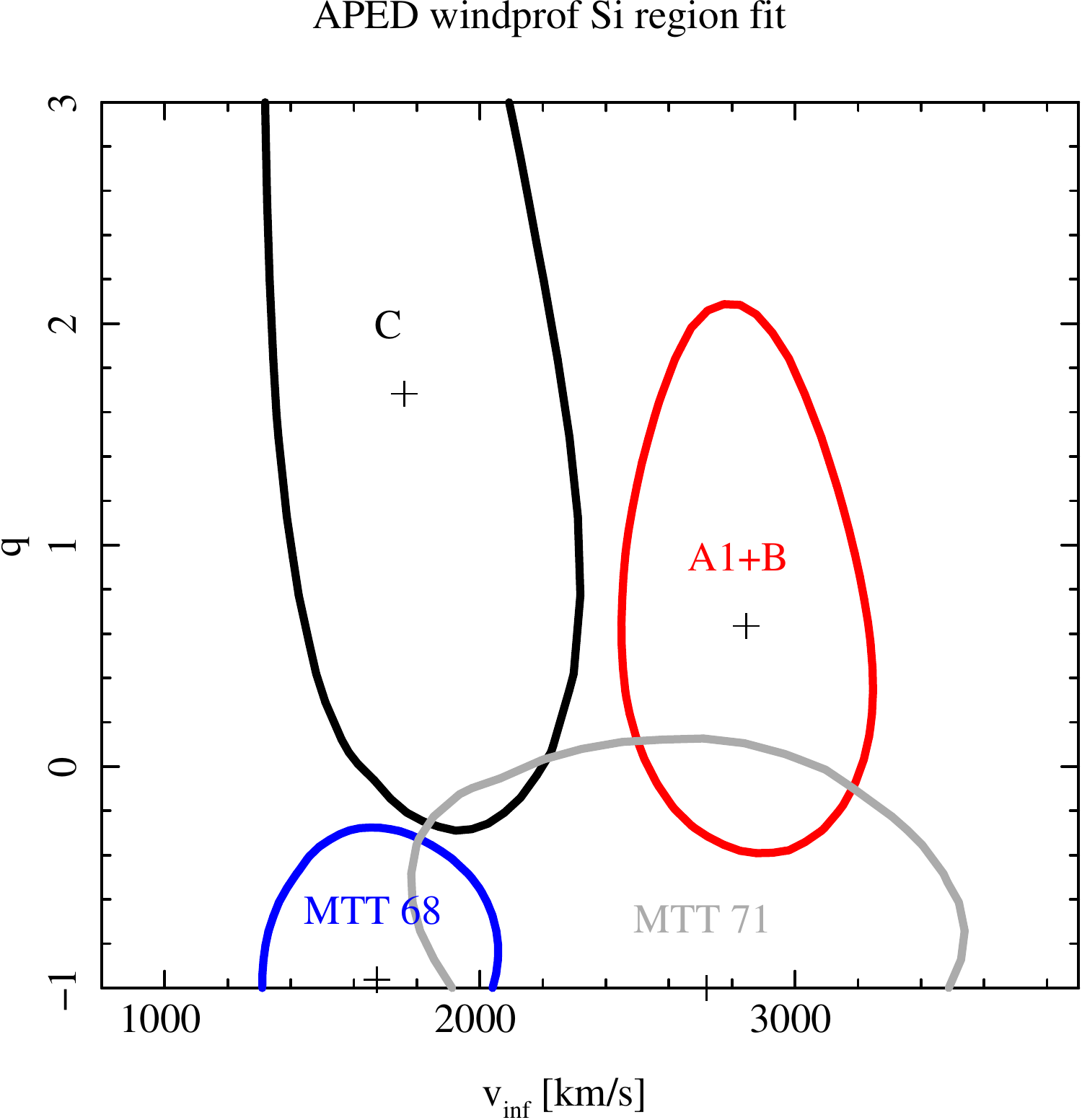}
  \caption{Asymptotic wind profile fits, confidence contours of
    $v_\infty$ and the shape paramter, $q$ ($90\%$ contours).  Fits
    were constrained to the \eli{Si}{14}--\eli{Si}{13} region.}
  \label{fig:confmapb}
\end{figure}

Figures~\ref{fig:confmapa}--\ref{fig:confmapb} show the confidence
contours (90\% limits) for the interesting shape parameters for the
Gaussian and wind-profile models respectively.  For the Gaussian,
Doppler-shifted fits, we corrected for our mean line-of-sight
velocities, as well as they are known (see Table~\ref{tbl:srcprops}),
and mark the expected velocity range of the single and double-lined
binary systems (A1 and C).  Best fits and corresponding $90\%$
errorbar values shown in
Figures~\ref{fig:confmapa}--\ref{fig:confmapb} are listed in
Table~\ref{tbl:linepar}. 

\begin{deluxetable}{ccccc} 
  \tablecaption{ {Line Fit Parameters} }
  \tabletypesize{\normalsize}
  \tablehead{
    \colhead{Object}&
    \colhead{$v_\infty$}&
    \colhead{$\sigma_{90}$\tablenotemark{a}}&
    \colhead{$FWHM$}&
    \colhead{$\sigma_{90}$}\\
    \label{tbl:linepar}
    &
    \multicolumn{2}{c}{($\kms$)}&
    \multicolumn{2}{c}{($\kms$)}
  }
  \startdata
  A1+B&     2845&   400&    2603&   932\\
  C&	      1760&   461&    1560&  1200\\
  MTT~68&     1672&   371&    2298&   684\\
  MTT~71&     2719&   870&    4117&  1945\\
  \enddata              
  \tablenotetext{a}{The approximate one-sided 90\% confidence limit,
    as measured from the 90\% contours shown in
    Figures~\ref{fig:confmapa}--\ref{fig:confmapb}.} 
\end{deluxetable}

\section{X-ray Light Curves}\label{sec:xrlc}

\begin{figure}[!htb]
  \centering\leavevmode
  \includegraphics*[width=0.45\columnwidth]{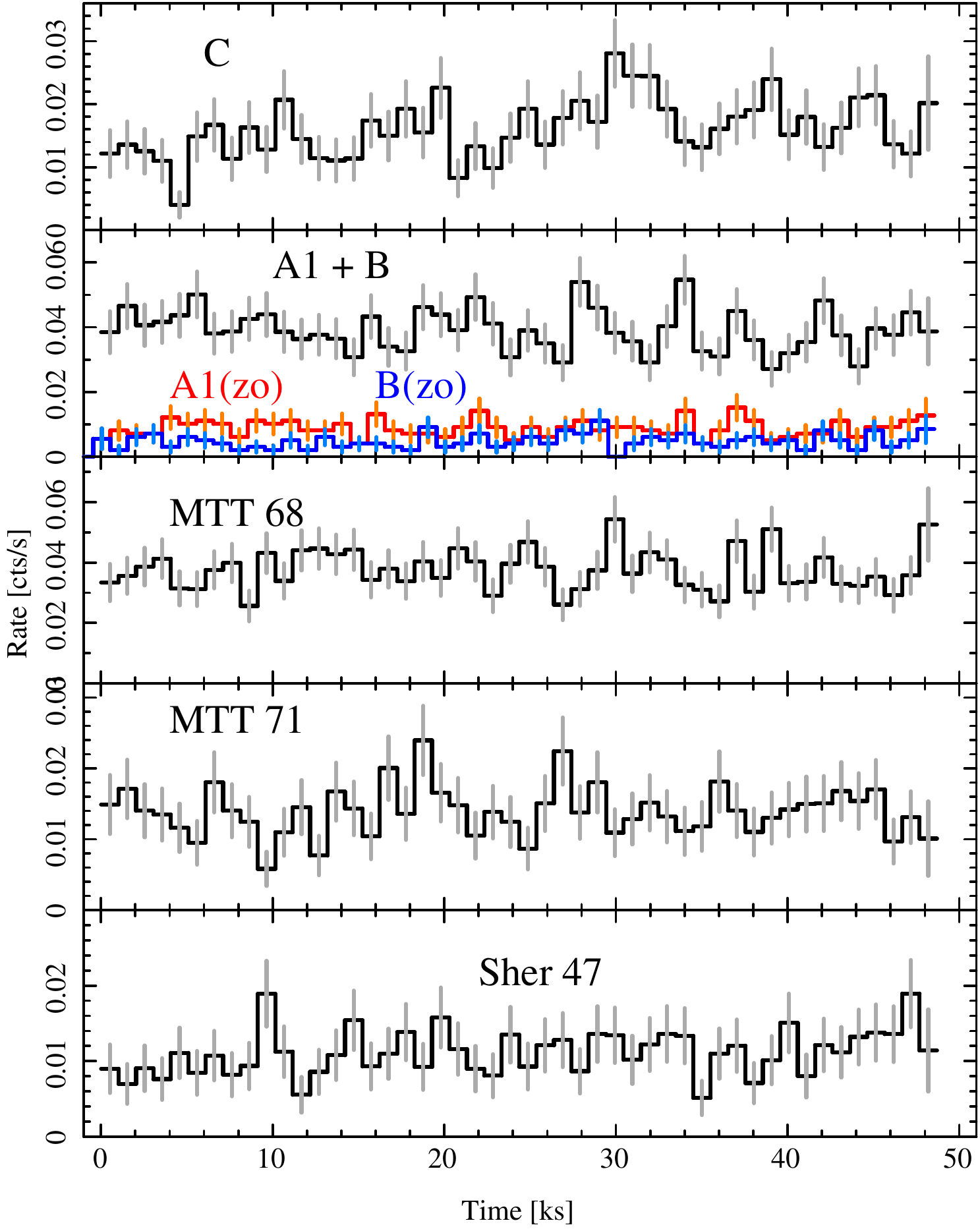}
  \caption{X-ray light curves of the selected stars with $1\ks$ bins.
    The black histogram in each panel shows the sum of the zeroth
    order and dispersed first orders count rates.  Since stellar
    components A1 and B are combined in the dispersed spectrum, we
    also show their individual zeroth order curves.  There is no
    obvious systematic or significant variability in any of the
    sources.  {One-sample Kolmogorov-Smirnov tests give
      probabilities of the curves being constant at their mean rate
      and normally distributed of 0.8 (C), 0.9 (A1+B), 0.9 (MTT~68),
      0.6 (MTT~71), and 0.7 (Sher~47).} }
  \label{fig:lc}
\end{figure}

We extracted light curves from the zeroth order and dispersed spectra
for each of the stars studied.  Figure~\ref{fig:lc} shows the count
rates. In the zeroth orders, we can isolate stars A1 and B, so we show
their zeroth orders individually, along with their sum and the
blended count rate in dispersed light.

\section{Discussion}\label{sec:disc}

X-ray line emission can be a sensitive probe of the structure and
energetics of stellar winds from massive stars.  The wind temperatures
and abundances can be determined from line and continuum emission.
Wind radial structure can be determined from He-like triplet ratios
which are strongly affected by UV photoexcitation.  Wind dynamics can
be revealed by X-ray line profiles.  It is important that we obtain
line profile information on WN stars to understand the wind structure.
While the central region of \ngc is crowded, and two of the components
(A1 and C) are binary systems, we have demonstrated that useful
information can be obtained with the \chan/\hetg.  Since our
observation was relatively short, we cannot determine the emission
line profiles' shapes in detail, but we can reliably measure centroids
and widths.  Though there is some unresolvable spatial overlap among
some sources, we can still obtain useful mean quantities.

\hdn-C is not a single star, but its optical spectrum is dominated by
the WN6h component, and a high mass ratio is implied
\citep{Schnurr:Casoli:al:2008}.  The relatively long period of 9 days
also means that the companion is likely a later-type star, not a
massive O-star.  Hence, we expect the WR star to dominate the emission
and we might not expect there to be a strong signature of colliding
winds.  The X-ray emission lines of \hdn-C are significantly
broadened, characteristic of $v_\infty \sim 1800 \kms$, which is
somewhat less than the value determined from UV lines (see
Tables~\ref{tbl:srcprops} and~\ref{tbl:linepar}).  The lines have a
significantly blueshifted centroid (see Figure~\ref{fig:confmapa}),
indicating a strong wind signature of absorption on the red wing.  The
shift is much larger than is expected from any Doppler effect from
binary orbital motion.

\hdn-A1 is a double-lined binary, and is spatially confused with
\hdn-B in the \hetg spectrum.  Fortunately, all the components have
the same spectral type of WN6h, so we can obtain mean parameters.
\hdn-A1 dominates, having a somewhat brighter zeroth order
(Table~\ref{tbl:rates}); the rates of A1 and B are consistent with 3
identical WN6h stars.  These stars have a very high $v_\infty$ of
about $3000\kms$, but are consistent with the UV measurements at their
90\% confidence levels.  The lines, like component C, also have a
significantly blueshifted centroid, larger than binary orbital motion
could cause.  The lines may have the ``fin'' shape characteristic of
the asymptotic flow of an optically thick wind, but this is
inconclusive and would require a much longer exposure to quantify.
{Some of the model profile mis-matches apparent in
  Figure~\ref{fig:sifits} may simply be due to the wrong detailed
  profile shape.  They may be more like the family of wind profiles
  derived by \cite{Owocki:Cohen:2001}.  We attempted parametric
  fitting (sum of Gaussians) of the \eli{Si}{13} region shown in
  Figure~\ref{fig:sifits}, but while the fit statistic can be reduced,
  resulting line ratios are non-physical (e.g.,
  forbidden-to-intercombination line ratios far above that allowed by
  standard plasma models).  It could be that better spectra will
  support source-model-independent measurement and confirm such
  anomalies, but for the present analysis, we prefer the plamsa-model
  based methods as a first characterization of these spectra. }

MTT~68 is a very early-type star of spectral type O2~If*
\citep{Roman-Lopes:2013b}.  It has an un-shifted but significantly
broadened profile, as determined from the Gaussian profile fit
(Figure~\ref{fig:confmapa}).  The ``fin'' profile fit is then forced
to have a low $q$ (more flat-topped) to make the line centered on its
rest wavelength; this fit serves to characterize the wind velocity
from the extent of the blue wing, to about $1700\kms$, but we prefer a
Gaussian (or more weakly skewed) profile.  This make the wind likely
different from the WNh stars' winds.

MTT~71 also has an early spectral type, O4~III.  It is fainter than
MTT~68, so not as well characterized.  We can confirm that its lines
are broadened, with $v_\infty\sim2700\kms$, though with large
uncertainty.  Whether the lines are shifted or non-Gaussian we cannot
reliably determine.

Sher~47 (O4~IV(f)), was prominent in the field and had a visible
dispersed spectrum.  However, its MEG spectrum was confused with that
of \hdn-A1+B, so could not be used.  There was not enough line signal
in the HEG spectrum to warrant detailed line fits.

We expect winds with embedded shocks to be relatively cool in X-rays
($kT \lesssim 1\kev$).  Due to the high line-of-sight absorption, we
cannot detect any very cool plasmas since the absorption hides the low
temperature, long-wavelength component (above about $12\mang$ or below
about $1\kev$).  From the presence of strong Si and Mg lines, we know
we have plasmas with $T\sim10\mk$.  From presence of S, possibly
\eli{Fe}{25}, and the continuum, we know we have very high
temperatures.  These stars all have a rather hot plasma component,
generally in excess of $50\mk$, with comparable or even greater
emission measure than the $\sim10\mk$ component
(Table~\ref{tbl:specpar}).  This high temperature is also obvious via
the continuum emission seen in the counts spectra in the $2$--$4\mang$
region (Figure~\ref{fig:summaryspec}).  For comparison, WR~6 also has
a high-temperature component of about $50\mk$, but it is has a much
lower weight relative to the lower temperature plasmas.  This is
suggestive of colliding winds, though we don't understand the origin
of the $50\mk$ plasma far out in the wind of the single WR~6.

Colliding wind binaries can produce high temperature plasmas with
quite diverse possible X-ray line profiles \citep{Henley:al:2003},
dependent upon the viewing geometry.  Emission line skewness depends
on the continuum optical depth and path length the wind collision
region.  Line emission can also have strong time dependence due to the
changing aspect, density, and wind velocity in the collision region
throughout the binary orbit.

It is likely that the X-ray emission from the stars we have
investigated includes emission from colliding winds.  Massive O and WR
stars are known to be X-ray sources even as single stars, so colliding
winds are an additional source of X-ray emission for these massive
stars.  Therefore it is difficult to determine if colliding winds are
contributing to the X-ray emission of a source unless the data cover
at least one full binary orbit, in which case there may be variability
in the light curve modulated by the orbit.  Our observation is only
about a half day out of the $3.77\days$ period for A1 and the
$8.9\days$ period of C, so we cannot verify colliding winds in these
objects.

We did examine the X-ray light curves (Figure~\ref{fig:lc}), and all
appear to be essentially constant. {According to one-sample KS-tests
assuming they are constant at their mean count rate, and with a
variance in accordance with their mean counts, they have a high
probability of being constant and normally distributed.}

Two WR stars known to have some X-ray emission due to colliding winds
(WR~140, \citet{pollock2005}; WR~25, \citet{pollock:corcoran:2006})
show strong variability in their X-ray light curves phased with their
highly eccentric binary orbits.  {These are long-period systems
  and certainly unlike A1, B, or C in many regards, but they are
  relevant in showing the dramatic effect of wind collisions on X-ray
  luminosity, and the dependence on binary separation.  Some WR stars
  with shorter periods are WR~46 ($P=0.3\days$), or WR~6
  ($P=3.8\days$). While WR~46 is variable in the UV, it is not
  convincingly variable in X-rays \citep{zhekov:2012}.  WR~6 is
  variable in X-rays, but erratically, with small amplitudes over a
  day ($\sim10\%$), up to a factor of 2 over a long term, but no
  convincing X-ray periodicity \citep{Ignace:al:2013}.}  Our derived
fluxes differ from those of \citet{Townsley:al:2014} by up to a factor
of two (see Tables~\ref{tbl:srcprops} and ~\ref{tbl:rates}), but their
determinations are uncertain because the \chan/\acis imaging-mode data
suffered from severe CCD count-rate saturation \citep[``pile-up'';
see][Table 4\footnote{\citet{Townsley:al:2014} source ``p1\_4736'' is
  MTT~71.}]{Townsley:al:2014}. It is not conclusive whether our fluxes
are indicative of long-term or orbital variability.  {While it
  is not unusual for stars like these to be constant in X-rays, longer
  term observations will be required to better detect and characterize
  any trends with orbital phase or epoch of observation.}

These stars have large X-ray luminosities, in excess of $10^{34}\lum$,
which is also high relative to their bolometric luminosities, having
$L_\mathrm{x}/L_\mathrm{bol} > 10^{-6}$.  This strongly suggests that
colliding winds are contributing to the X-ray emission of these stars.
Components A1 and C are known to be binaries.  The extreme stellar
density of the region implies a high likelihood of multiplicity for
all of these stars.  In general, binary WR stars are more
X-ray-luminous than single WR stars, having
$L_\mathrm{x}/L_\mathrm{bol} > 10^{-7}$ \citep{stevens:al:2002}. WR
binaries have a range of $\log(L_{\mathrm{x}}/L_{\mathrm{bol}}$) of
$-7.3$ to $-4.0$ \citep{gagne:al:2012}.  Specifically, the two
previously mentioned known colliding wind binaries, WR~140 and WR~25,
have $\log(L_{\mathrm{x}}/L_{\mathrm{bol}})$ of $-4.8$ and $-5.5$,
respectively.  For comparison, the single Wolf-Rayet star, WR~6 (type
WN4) has a luminosity 10--20 times lower than the stars studied here,
and has $L_\mathrm{x}/L_\mathrm{bol} < 10^{-6}$
\citep{Huenemoerder:al:2015,
  Oskinova:al:2012}. Table~\ref{tbl:specpar} shows the range of $-6.0$
to $-4.8$, which is well above the norm for single WR stars.  {The
X-ray luminosities of the \ngc stars also exceed the binary WR
stars in the sample studied by \citet{zhekov:2012}, where only WR~148
approaches $L_\mathrm{x}\sim10^{34}\lum$.}

The O-type stars in our sample, MTT~68, MTT~71, and Sher~47 (see
Table~\ref{tbl:srcprops}) are also highly luminous for their
classes. Single O-stars also have $L_x/L_\mathrm{bol} \sim 10^{-7}$
\citep{naze:broos:al:2011}.  In a review of binary colliding winds,
\citet{rauw:naze:2016}, from a variety of X-ray studies, conclude that
X-ray bright colliding wind binaries are relatively rare; being in a
binary does not necessarily confer high relative X-ray luminosity.
Two early O-type stars in the core of \ngc, A2 and A3, have not been
detected in X-rays though their bolometric luminosities are similar to
the detected O-stars; perhaps they are ``normal'', single systems.
Whether the X-ray bright O-type systems in \ngc are binary remains to
be determined.

\medskip
\section{Conclusions}\label{sec:conc}

The relatively short \chan/\hetg exposure has shown the feasibility of
study of this cluster at high X-ray spectral resolution.  For the
brighter components which are saturated in ACIS imaging observations,
we have provided more reliable flux measurements. We have confirmed
the high terminal wind velocities in X-rays for \hdn-A1, B, and C, and
we have provided the first empirical estimates of the terminal
velocities of MTT~68 and MTT~71 from X-ray line profiles.  
The high plasma temperatures and the very high values of
$L_{\mathrm{x}}/L_{\mathrm{bol}}$ strongly suggest a colliding winds
origin for their X-rays.

Further studies of \ngc stars in high resolution X-rays are needed to
refine the characteristics of their winds and the origin of their very
high X-ray luminosities.  Higher signal-to-noise ratios are needed to
determine emission line shapes in detail, as these are critical for
measurement of the terminal velocity and of the optical depths in
their winds.  Higher signal is also needed to measure He-like line
ratios of \eli{Mg}{11} and \eli{Si}{13}, since these are sensitive to
the UV radiation field and thereby serve as a proxy for distance of
formation from the photospheres.  Longer term X-ray studies are needed
in order to search for variability expected from colliding wind
binaries.  Two of the systems, A1 and C, are known to be binaries and
have relatively short periods, making them amenable to X-ray
variability studies.  Whether the other stars in this study are
binaries remains to be determined and is important for interpretation
of their properties.

%
%
%


\acknowledgements Acknowledgements: Support for this work was provided
by NASA through the Smithsonian Astrophysical Observatory (SAO)
contract SV3-73016 to MIT for Support of the Chandra X-Ray Center
(CXC) and Science Instruments. CXC is operated by SAO for and on
behalf of NASA under contract NAS8-03060.  We thank Prof.\ Claude
Canizares for allocation of observing time for this observation from
the \hetg Guaranteed Time Observation Program.

\facility{ CXO (HETG/ACIS) }

\software{MARX \citep{Davis:al:MARX:2012},  CIAO \citep{CIAO:2006},  ISIS \citep{Houck:00}}


%

\input{ms.bbl}

\end{document}

%% file: local_defs.tex
\newcommand{\mone}  {^{-1}}
\newcommand{\mtwo}  {^{-2}}
\newcommand{\mthree}{^{-3}}
\newcommand{\mfour} {^{-4}}

\newcommand{\mang}  {\,\mathrm{{\mbox{\AA}}}\xspace}
\newcommand{\mmang}  {\,\mathrm{{m\mbox{\AA}}}\xspace}

\newcommand{\cmmtwo}{\,\mathrm{cm\mtwo}}

\newcommand{\cmmthree}{\,\mathrm{cm\mthree}}

\newcommand{\cts}   {\,\mathrm{cts\,s\mone}}

\newcommand{\days}   {\,\mathrm{d}}
\newcommand{\eflux} {\,\mathrm{ergs\,cm\mtwo\,s\mone}}

\newcommand{\kev}   {\,\mathrm{keV}}
\newcommand{\kpc}   {\,\mathrm{kpc}}

\newcommand{\kms}   {\,\mathrm{km\,s\mone}}
\newcommand{\ks}    {\,\mathrm{ks}}
\newcommand{\pc}    {\,\mathrm{pc}}
\newcommand{\lum}   {\,\mathrm{ergs\,s\mone}}
\newcommand{\mk}    {\,\mathrm{MK}}

\newcommand{\acis}  {{ACIS}\xspace}

\newcommand{\arf}   {{ARF}\xspace}

\newcommand{\chan}  {{\it Chandra}\xspace}

\newcommand{\ciao}  {{CIAO}\xspace}

\newcommand{\heg}   {{HEG}\xspace}
\newcommand{\hetgs} {{HETGS}\xspace}
\newcommand{\hetg}  {{HETG}\xspace}

\newcommand{\meg}   {{MEG}\xspace}

\newcommand{\rgs}   {{RGS}\xspace}
\newcommand{\rmf}   {{RMF}\xspace}

\newcommand{\xmm}   {{\it XMM-Newton}\xspace}

\newcounter{ion} \newcommand{\eli}[2]{\setcounter{ion}{#2}#1{~\sc\roman{ion}}}

\newcommand{\ngc}{{NGC~3603}\xspace}
\newcommand{\hdn}{{HD~97950}\xspace}